\documentclass[twocolumn,aps,nofootinbib,showpacs,amsmath,amssymb]{revtex4}
\usepackage{amsfonts,amssymb,amscd,amsmath}
\usepackage{graphicx}
\usepackage{subfigure}
\usepackage{bm}
\graphicspath{{figures/}}
\def\im{{\rm Im}}

\def\gsim{\mathrel{\rlap{\lower4pt\hbox{\hskip1pt$\sim$}}
 \raise1pt\hbox{$>$}}}

 \newcommand\la{\langle}
 \newcommand\ra{\rangle}
 \newcommand\beq{\begin{equation}}
 
 \newcommand\eeq{\end{equation}}
 \newcommand\beqn{\begin{eqnarray}}
 \newcommand\eeqn{\end{eqnarray}}

\def\fm{\,\mbox{fm}}
\def\GeV{\,\mbox{GeV}}
\def\TeV{\,\mbox{TeV}}
\def\lsim{\mathrel{\rlap{\lower4pt\hbox{\hskip1pt$\sim$}}
    \raise1pt\hbox{$<$}}}         %less than or approx. symbol
\def\gsim{\mathrel{\rlap{\lower4pt\hbox{\hskip1pt$\sim$}}
    \raise1pt\hbox{$>$}}}         %greater than or approx. symbol

\def\fm{\,\mbox{fm}}
\def\GeV{\,\mbox{GeV}}

\begin{document}
%%%%%%%%%%%%%%%%%%

%%%%%%%%%%%%%%%%%%%%%%%%%%%%%%%%%%%%%%%%%%
\title{
Flavor-independent yield of high-\boldmath$p_T$ hadrons from nuclear collisions
}
%%%%%%%%%%%%%%%%%%%%%%%%%%%%%%%%%%%%%%%%%%%%%%%%%%%%%%%%%%%%%%%

\author{B. Z. Kopeliovich$^1$}
\email{boris.kopeliovich@usm.cl}

\author{J. Nemchik$^{2,3}$}
\email{jan.nemcik@fjfi.cvut.cz}
%\email{nemcik@saske.sk}

\vspace*{0.9cm}

\affiliation{
\vspace*{0.3cm}
$^1$
Departamento de F\'{\i}sica,
Universidad T\'ecnica Federico Santa Mar\'{\i}a,
Avenida Espa\~na 1680, Valpara\'iso, Chile}
\affiliation{
$^2$
Czech Technical University in Prague, FNSPE, B\v rehov\'a 7, 11519
Prague, Czech Republic}
\affiliation{
$^3$
Institute of Experimental Physics SAS, Watsonova 47, 04001 Ko\v sice, Slovakia
\vspace*{0.3cm}
}

\vspace*{0.3cm}
%%%%%%%%%%%%%%%%%%%%%%%%%%%%%%%%%%%%%%%%%%%%%%%%%%%%%%%%%%%%%%%

%
%
%
%%%%%%%%%%%%%%%%%
\begin{abstract}
%%%%%%%%%%%%%%%%%
\vspace*{0.7cm}
Data on high-$p_T$ hadron production in heavy ion collisions at Feynman $x_F=0$
indicate at universality of the observed nuclear suppression. Our analysis of the 
production mechanisms demonstrates important role of the color transparency effects which make  
the survival probability of a quark-antiquark dipole independent of the quark flavor, 
provided that the hadron wave function is formed outside the medium. 
The latter condition imposes restrictions on the range of $p_T$, 
which should be sufficiently high to make the nuclear suppression universal. 
We also found that the in-medium broadening rate $\hat q$  
(frequently called transport coefficient) significantly depends 
on the quark flavor, diminishing for heavy quarks.
%
%%%%%%%%%%%%%%%
\end{abstract}
%%%%%%%%%%%%%%%
%
%
%

%%%%%%%%%%%%%%%%%%%%%%%%%%%%%%%%%%%%%%%%%%%%%%%%%%%%%%%%%%%%%%%

\date{\today}


\maketitle

%
%
%
%%%%%%%%%%%%%%%%%%%%%%%%%%
\section{Introduction}
%\label{intro}
%%%%%%%%%%%%%%%%%%%%%%%%%%
%
%
%

Measurements of high-$p_T$ hadron production in heavy ion collisions (HICs)
from the ALICE experiment \cite{alice-u14,alice-u16}
revealed a remarkable universality of the nuclear suppression factors $R_{AA}$ 
for different species of produced hadrons, such as pions, kaons and protons.
Besides, the ATLAS data on prompt production of $J/\Psi$ \cite{atlas18-prompt-psi}
demonstrate a similar attenuation as was measured
for the light hadrons.
Such an observation can be hardly described by models based on the energy-loss 
scenario (see Ref.~\cite{miklos}, for example)
where only induced energy loss by a parton propagating through the dense medium
represent the main reason for the high-$p_T$ hadron suppression.
Here the hadronization process, closely connected with a different radiation
by light and heavy quarks, cannot lead to similar magnitudes of $R_{AA}$
for the light and heavy flavored particles.
Besides, different mechanisms for production of mesons and baryons 
should naturally exclude the universality of $R_{AA}$.
In the present paper we demonstrate an alternative scenario of the in-medium
hadronization process with a proven shortness of the hadronization length
\cite{green,my-alice,simple,within}.
Then the main source of suppression of the production rate comes from  
the survival probability of the produced colorless dipoles 
propagating through the medium. 
Whereas at small and medium values of $p_T$ the dipole size 
is controlled by quark masses, 
in the region of sufficiently high $p_T$ it decreases with $p_T$ 
and the quark mass does not play any role.
This is a typical manifestation of the color transparency (CT) effect
\cite{zkl,bertsch}, 
which describes the flavor independent attenuation of propagating dipoles
in the medium and thus allows to explain the observed universality
in production of different hadrons.

The paper is organized as follows.
In the next Sec.~\ref{lp-vacuum} we prove briefly that gluon radiation 
ceases shortly after a hard collision due to color neutralization 
and production of a colorless dipole. 
We demonstrate that the duration of this stage of hadronization, 
accompanied by the vacuum energy loss, is very short and decreases with
the quark mass and $p_T$.
In Sec.~\ref{mesons} we present model predictions for 
the nuclear modification factor $R_{AA}$ as function of $p_T$
for production of various hadronic species in HICs. 
We prove that the dominant source of suppression turns out to be related to 
propagation in the medium of high-$p_T$ hadrons during formation of 
their wave functions.
The regime of universal suppression onsets at sufficiently large  
$p_T>p_T|_{min}^h$.
The corresponding values of $p_T|_{min}^h$ are evaluated for pions, kaons, 
$J/\psi$, $\Upsilon$, as well as open heavy flavored $D$ and $B$ mesons. 
We demonstrate that model calculations are in a good agreement with available data.
The results of the paper are summarized and discussed in Sec.~\ref{conclusions}.

%
%
%
%%%%%%%%%%%%%%%%%%%%%%%%%%%%%%%%%%%%%%%%%%%
\section{Production length in vacuum
%==================
\label{lp-vacuum}
%==================
}
%%%%%%%%%%%%%%%%%%%%%%%%%%%%%%%%%%%%%%%%%%%
%
%
%

The popular scenario for quenching of high-$p_T$ hadrons in heavy ion collisions 
is based on the medium-induced energy loss model (see e.g.  Ref.~\cite{miklos}).
The main assumption of the model, which has never been justified, is a long hadronization length.
The process is assumed to end up with production of the detected hadron 
well outside of the dense medium created in HICs. 
This assumption was challenged in Refs.~\cite{similar,my-alice,green} with a detailed study of the hadronization dynamics, which led to a conclusion that hadronization length $L_p$ is very short, 
of the order of $1\div 2\,\fm$. 
The main reason for that is the intensive gluon radiation by a highly virtual parton 
originated from a high-$p_T$ hard reaction. 
Here the important issue is energy conservation.
Namely, a parton during the hadronization process may dissipate so much energy that
cannot produce a hadron with large light-front (LF) momentum fraction $z$.
Let us remind the main observations of that consideration:
\\
%
%%%%%

(i)
%%%%%
The mean hard scale of the process $\bar Q=\sqrt{q_T^2+m^2}$ and the jet energy coincide. 
Here $q_T$ and $m$ are the transverse momentum and mass of the initially produced parton;
\\
%
%%%%%

(ii) 
%%%%%
A parton originated from a hard process is lacking a part of its color field 
with transverse frequencies $k_T$ up to the scale of the process $k_T\lesssim \bar Q$. 
It regenerates the stripped off field by radiating gluons. 
The perturbative spectrum of radiation reads,
%
%----------------------------------------------
\beq
\frac{dn_g}{dx\,dk_{T}^2} =
\frac{2\alpha_s(k_{T}^2)}{3\pi\,x}\,
\frac{k_{T}^2[1+(1-x)^2]}{[k_{T}^2+x^2m_Q^2]^2}\,,
%===========
\label{145}
%===========
\eeq
%---------------------------------------------
%
where $x$ is the fractional LF momentum taken away by the radiated gluon.
The gluons are radiated, i.e. get to mass shell, when lose the coherence 
with the source quark due to large a phase shift between the Fock components 
of the quark, $|q\ra$ and $|qg\ra$. 
So the coherence length for gluon radiation is given by,
%
%---------------------------------------------
\beq
L^g_c=\frac{2E\,x(1-x)}{k_{T}^2+x^2\,m_q^2}\,.
%===========
\label{140}
%===========
\eeq
%---------------------------------------------
%
Here $E=E_T$ is the energy of the quark in the reference frame, 
where it has no longitudinal momentum.

As far as the radiation length is known, one can evaluate how much energy 
is radiated along the path length $L$,
%
%-----------------------------------------------------
\beq
\Delta E_{rad}(L) =
\int\limits_{\lambda^2}^{\bar Q^2}
dk_{T}^2\int\limits_0^1 dx\,\omega\,
\frac{dn_g}{dx\,dk_{T}^2}\,
\Theta(L-L^g_c)\,,
%===========
\label{130}
%===========
\eeq
%-----------------------------------------------------
%
where $\omega = x E$ is the gluon energy
and $\lambda = 0.2\,\GeV$ is the soft cut-off parameter.

\begin{figure}[hbt]
    \includegraphics[height=8.9cm]{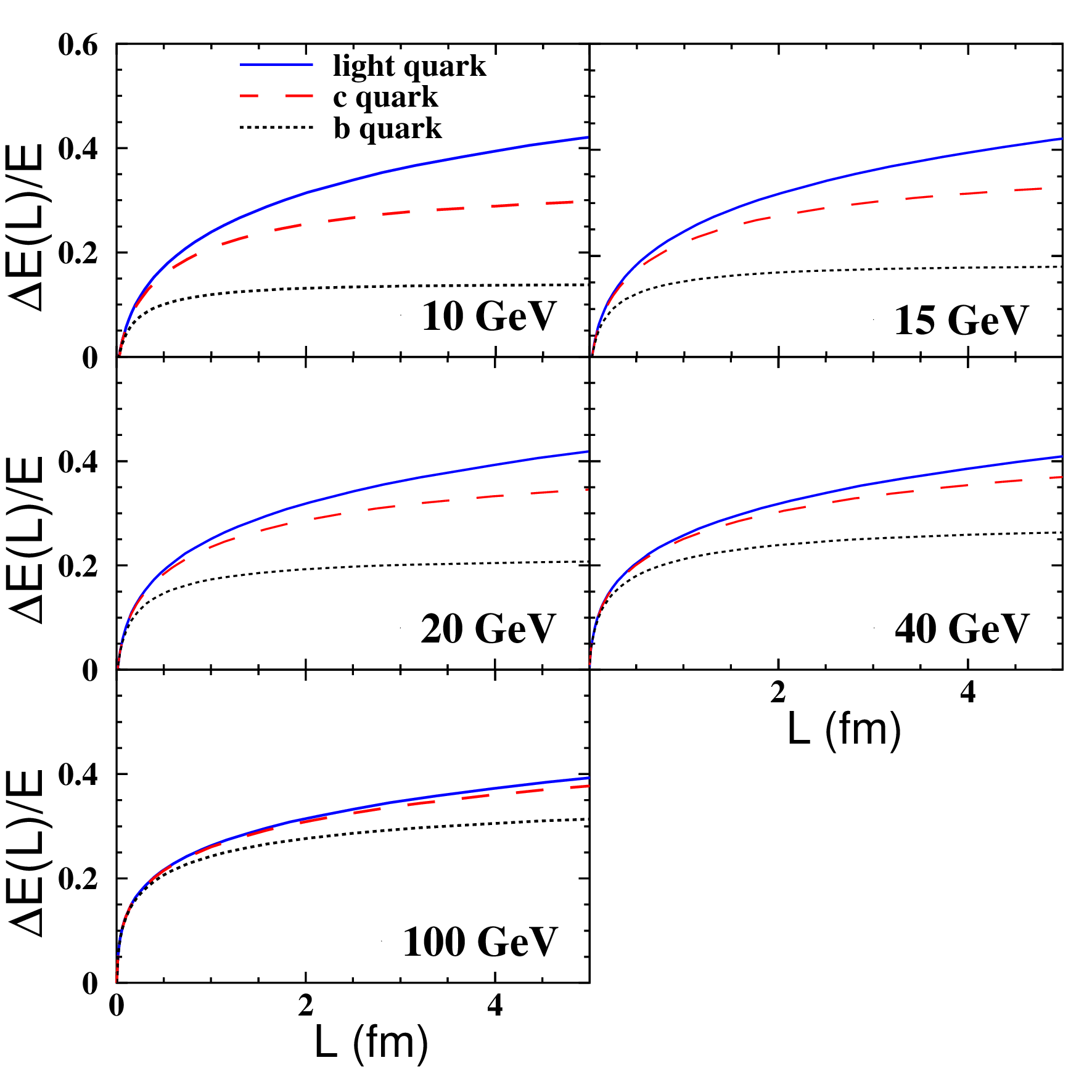}
    \caption{ 
         %===================
         \label{fig:eloss-qn}
         %===================
         (color online)
         Fractional radiational energy loss in vacuum by light, c and b quarks 
         as function of path length for  
         different initial quark energies $E=\sqrt{p_T^2+m_q^2}$.
         }
\end{figure}

Now one can trace how much energy is radiated by a high-$q_T$ quark as function of $L$.
Some examples are depicted in Fig.~\ref{fig:eloss-qn} for different quark energies.
%
%
%                         FIG.1
%%%%%%%%%%%%%%%%%%%%%%%%%%%%%%%%%%%%%%%%%%%%%%%%%%%%%%%%%%%%%%%%%%%%%%
%%%%%%%%%%%%%%%%%%%%%%%%%%%%%%%%%%%%%%%%%%%%%%%%%%%%%%%%%%%%%%%%%%%%%%%
%
%
%
One can see that a substantial fraction of the initial parton energy 
is radiated at a short path length of the order of fm. 
At longer distances radiation ceases;
\\
%
%%%%%%

(iii) 
%%%%%%
Fig.~\ref{fig:eloss-qn} demonstrates that the fraction $\Delta E_{rad}$ 
of the total radiated energy depends on the quark mass as was analyzed in Refs.~
%----------------------
\cite{knp,within,similar}.
%----------------------
The heaver the quark is, the smaller fraction of its energy can be radiated.
This is a result of the \textit{dead cone effect} \cite{dk} 
in accordance with Eq.~(\ref{145}).

Fig.~\ref{fig:eloss-qn} also shows a substantial difference between 
radiational dissipation of energy by heavy and light quarks 
\cite{hm17,hm18,universal19,hq19a,hq19b,hm23}.
One can see that radiation by heavy quarks ceases shortly.
While heavy quarks radiate only a small fraction of 
their initial energy, $\Delta z=\Delta E_{rad}/E$, light quarks
lose for radiation most of the initial energy;
\\
%
%%%%%

(iv)
%%%%%
The $p_T$-dependence of the hadron production rate is given by convolution of the
$q_T$ dependent parton production cross section and the fragmentation function 
of a quark to hadron $D_{q/h}(z)$,
where $z=p_T/q_T$,
%
%-------------------------------------------------
\beqn
\frac{d^2\sigma_{pp\to hX}}{d^2p_T}
=\frac{1}{2\pi p_T E_T}
\int d^2q_{T}
\frac{d^2\sigma_{pp\to qX}}{d^2q_{T}}\,
z\,D_{q/h}(z)\,.
%===========
\label{160}
%===========
\eeqn
%------------------------------------------------
%
Extension to a gluon fragmentation is straightforward.

In the convolution Eq.~(\ref{160}) the mean value of $z\sim 0.6-0.7$ 
depending on the collision energy and hadron transverse momentum $p_T$ 
as was evaluated in Ref.~\cite{green}.
The corresponding calculations were based on the 
model-independent evaluation of the time of hadronization,
which stops with the neutralization of the leading quark color by picking up
an antiquark. Here we adopted the Berger model \cite{berger},
which assumes equal sharing of the dipole LF momentum between $q$ and $\bar q$.

%
%
%
%%%%%%%%%%%%%%%%%%%%%%%%%%%%%%%%%%%%%%
\section{Production of high-$p_T$ mesons 
%===============
\label{mesons}
%===============
}
%%%%%%%%%%%%%%%%%%%%%%%%%%%%%%%%%%%%%%
%
%
%

%------------------------------------------------
%------------------------------------------------
\subsection{Light mesons produced in HICs
%===============
\label{lm-hics}
%===============
}
%------------------------------------------------
%------------------------------------------------

Besides of energy dissipation in vacuum as was described 
in the previous Sec.~\ref{lp-vacuum}, an additional
medium-induced energy loss makes the production length $L_p$
even shorter.

In this section we consider non-strange $|q\bar q\ra$, as well as strange $|s\bar q\ra$ mesons. 

Shortness of the production length implies factorization of short 
and long distances in the cross section of high-$p_T$ hadron production 
in collisions of nuclei $AA$ (taken identical for the sake of simplicity) 
with relative impact parameter $\vec b$. 
Then the ratio of the $AA$-to-$NN$ cross sections, properly normalized, 
can be represented as an integral over the transverse overlap of the colliding nuclei,
%
%------------------------------------------------------------
\beqn
R_{AA}(p_T,b)
&=& 
\frac{1}{T_{AA}(b)}
\int d^2\tau\, T_A(\vec \tau)\, T_A(\vec b-\vec \tau)
\nonumber\\
&\times&
 \int\limits_0^{2\pi} \frac{d\phi}{2\pi}\,
 S_{Q\bar q}(\vec b, \vec \tau, E_T,\phi)\,,
%============
\label{215}
%===========
\eeqn
%--------------------------------------------------------------
%
where 
$\vec\tau$ is the impact parameter of the hard parton-parton collision; 
$T_{AA}(b) = \int d^2\tau\, T_A(\vec \tau)\, T_A(\vec b-\vec \tau)$ 
is the nuclear thickness function, 
$T_A(\tau)=\int_{-\infty}^\infty dz'\,\rho_A(\tau,z')$, 
is given by the integral over longitudinal coordinate $z'$ of 
the nuclear density $\rho_A(\vec r)$; 
$\phi$ is the azimuthal angle between the quark trajectory and reaction plane.

The nuclear suppression factor $S_{Q\bar q}(\vec b,\vec\tau,E_T,\phi)$ 
in Eq.~(\ref{215}) is the survival probability of the $Q\bar q$ dipole 
produced in the hard collision with starting transverse separation $r_1$, 
and propagating through the medium without inelastic collisions.
It has the following form,

%
%----------------------------------------------------------------
\begin{widetext}
\beqn
S_{Q\bar q}(\vec{b},\vec\tau,E_T,\phi) 
=
%\hspace*{6.0cm} \nonumber\\
\frac{
\left|
\int\limits_0^1d\alpha\int d^2r_1\, d^2r_2\,
\Psi_h^\dagger(\vec r_2,\alpha)\,
G_{Q\bar q}(\vec{b},\vec{\tau},E_T,\phi \,|\, L_1,\vec r_1;L_2,\vec r_2)\,
\Psi_{in}(\vec r_1,\alpha)\right|^2
}{
\left|\int\limits_0^1d\alpha\int d^2r_1\, d^2r_2\,
\Psi_h^\dagger(\vec r_2,\alpha)\,
%G_{\bar qq}(l_1,\vec r_1;l_2,\vec r_2)
\Psi_{in}(\vec r_1,\alpha)\right|^2}\,.
%===========
\label{270}
%===========
\eeqn
%---------------------------------------------------------------
%
\end{widetext}
It is subject to color transparency effects:  
the smaller is the dipole, the higher is its chance to survive in the medium. 
The most effective way of calculation of the factor $S_{Q\bar q}$ 
is summing up all trajectories of $Q$ and $\bar q$. 
This leads to the path integral formalism, which gives rise to the factor  
$G_{Q\bar q}(\vec{b},\vec{\tau},E_T,\phi \,|\, L_1,\vec r_1;L_2,\vec r_2)$ in Eq.~(\ref{270}).
The Green function $G_{Q\bar q}(L_1,\vec r_1;L_2,\vec r_2)$ describes evolution
of a dipole starting from transverse separation $\vec r_1$ 
at the longitudinal coordinate  $L_1$ up to size $\vec r_2$ 
at the longitudinal distance $L_2$, in accordance with 
the two-dimensional Schr\"odinger equation
%-----------------------------------
\cite{kst1,kst2,krt,zakharov,green},
%-----------------------------------
%
%---------------------------------------------------------------
\begin{widetext}
\beqn 
i\frac{d}{dL_2}\,G_{Q\bar q}(L_1,\vec r_1;L_2,\vec r_2)=
\left[\frac{\mu^2 - 
\Delta_{r_2}}{2\,E_T\,\alpha\,(1-\alpha)}\,
-\,V_{Q\bar q}(L_2,\vec r_2)\right]\,
G_{Q\bar q}(L_1,\vec r_1;L_2,\vec r_2)\,,
%===========
\label{280}
%===========
\eeqn
\end{widetext}
%---------------------------------------------------------------
%
with the boundary conditions
%
%---------------------------------------------------------------
\beqn
G_{Q\bar q}(L_1,\vec r_1;L_2,\vec r_2)\Bigr|_{L_1=L_2} 
&=&
\delta^{(2)}(\vec r_2-\vec r_1);
\nonumber\\
G_{Q\bar q}(L_1,\vec r_1;L_2,\vec r_2)\Bigr|_{L_1>L_2} 
&=& 
0\,.
%===========
\label{290}
%===========
\eeqn
%--------------------------------------------------------------
%

In Eq.~(\ref{280}) $\mu^2=(1-\alpha)\,m_q^2 +\alpha\,m_Q^2$; 
$E_T=\sqrt{p_T^2+ (m_q+m_Q)^2}$ is the energy of the $Q\bar q$ dipole.

The imaginary part of the LF potential $V_{Q\bar q}(L_2,\vec{r}_2)$
in Eq.~(\ref{280}) is responsible for absorption of the $Q\bar q$ dipole
in the medium and can be expressed through the broadening rate $\hat q$ 
(usually called transport coefficient) as follows
%------------
\cite{green},
%------------
%
%---------------------------------------------------
\beq
\im V_{Q\bar q}(L,\vec r) = 
- \frac{1}{4}\,\hat q(L)\,r^2\,.
%===========
\label{300}
%===========
\eeq
%--------------------------------------------------
%
It varies with transverse coordinates and time.
We rely on the popular model from Ref.
%----------------
\cite{hat-q},
%----------------
%
%
%---------------------------------------------------------------------
\beqn
\hat q(t,\vec b,\vec\tau)=\frac{\hat q_0\,t_0}{t}\,
\frac{n_{part}(\vec b,\vec\tau + t\,\vec{p}_T/p_T)}{n_{part}(0,0)}\,
\Theta(t-t_0)\,,
%===========
\label{170}
%===========
\eeqn
%---------------------------------------------------------------------
%
where $\hat q$ is proportional to the number of participants $n_{part}$
depending on the transverse coordinate $\vec\tau$ 
and impact parameter $\vec b$ of the collision.
Its  maximal value $\hat q_0$ is reached for the central collision at $b = \tau = 0$ 
and at $t=t_0$, which is the time scale for medium formation, $t_0\sim 1\,\fm$.
The value of $\hat q_0$ is difficult to predict reliably, so it is the fitted
parameter in the analysis of data. 
The falling time dependence is 
associated 
with longitudinal expansion of the medium. 
It is related to the path length of the dipole as
$L = t\,\sqrt{1 - (m_q+m_Q)^2/E_T^2}$.

The real part of the light-front potential $V_{Q\bar q}$ in Eq.~(\ref{280})
describes the nonperturbative interaction between $Q$ and $\bar{q}$
in the dipole as was discussed in Refs.~
%--------------
\cite{kst2,VM}.
%--------------
Here we treat the $Q\bar q$ system as free non-interacting
partons, like in Ref.~
%------------
\cite{green},
%------------
i.e. assuming maximal effects of absorption.
We checked that the real part of potential does not affect much 
the dipole evolution during the initial perturbative stage of its
development.

The dependence of the hadron wave function $\Psi_h(\vec r,\alpha)$ in Eq.~(\ref{270}) 
on the transverse $Q\bar q$ dipole separation $r$ and LF momentum fraction $\alpha$ 
is taken according to the prescription for the Lorentz boost 
from the $S$-wave state in the hadronic rest frame to the LF frame used in Refs.~
%------------------------------------
\cite{jan-vm-97,ihkt-wf,jan-19},
%-------------------------------------
%
%-------------------------------------------------------------------------
\beqn
\Psi_h (\vec r,\alpha) 
&=& 
N\,
%(2\pi R_h^2)^{3/2}\, 
\frac{4\alpha(1-\alpha)}{R_h^2}
%=============
\label{305}\\ 
%=============
&\times&
\exp\left[- \frac{2 \alpha(1-\alpha)\,r^2}{R_h^2}
- \frac{(1-2\alpha)^2\,\mu^2\,R_h^2}{8\alpha(1-\alpha)}\right]\,,
\nonumber
\eeqn
%------------------------------------------------------------------------
%
where $R_h^2=8/3\,\la r_{ch}^2\ra_h$, and the wave function squared is normalized to unity,
$\int d^2 r\,d\alpha\, |\psi_h(\vec r,\alpha)|^2 = 1$.

A high-$p_T$ $Q\bar q$ dipole is produced with a small initial size, 
controlled by the hard scale of the process,  $r\sim 1/{\bar Q}\sim 1/E_T$. 
Therefore we take in Eq.~(\ref{270}) the initial dipole size distribution function in the form,
%
%--------------------------------------
\beq
\Psi_{in}(r)=\exp\, \bigl [- \frac{1}{2}\,E_T^2\,r^2\,\bigr ]\,.
%===========
\label{307}
%===========
\eeq
%--------------------------------------
%

Due to color transparency the survival probability of a dipole
propagating through the medium rises when the dipole size shrinks. 
As function of $p_T$ the initial dipole separation, $r_0\sim 1/E_T$, 
becomes smaller, and the dipole size expands quickly, proportionally to the path length.
At longer distances, however, Lorentz time dilation slows down the dipole expansion.
The dipole size expansion at the early stage is so fast that 
the initial dipole very small size $r_0$ is quickly "forgotten" 
and is not important anymore at longer path length 
%--------------------
\cite{my-alice,green},
%---------------------
%
%----------------------------------------------
\beq
\la r^2(t)\ra
=\frac{2\,t}{\alpha(1-\alpha)\,E_T}+r_0^2\,.
%===========
\label{202}
%===========
\eeq
%----------------------------------------------
%
This naturally explains the rising $p_T$ dependence of $R_{AA}(p_T)$ clearly seen in data.

Notice that Eq.~(\ref{202}) presents an oversimplified description 
of the dipole evolution, which allows to understand a qualitative 
features of the dipole expansion. In what follows we rely on the 
accurate quantum-mechanical description based on the path-integral technique.

\begin{figure}[hbt]
    \hspace*{-0.3cm}
    \includegraphics[height=8.7cm]{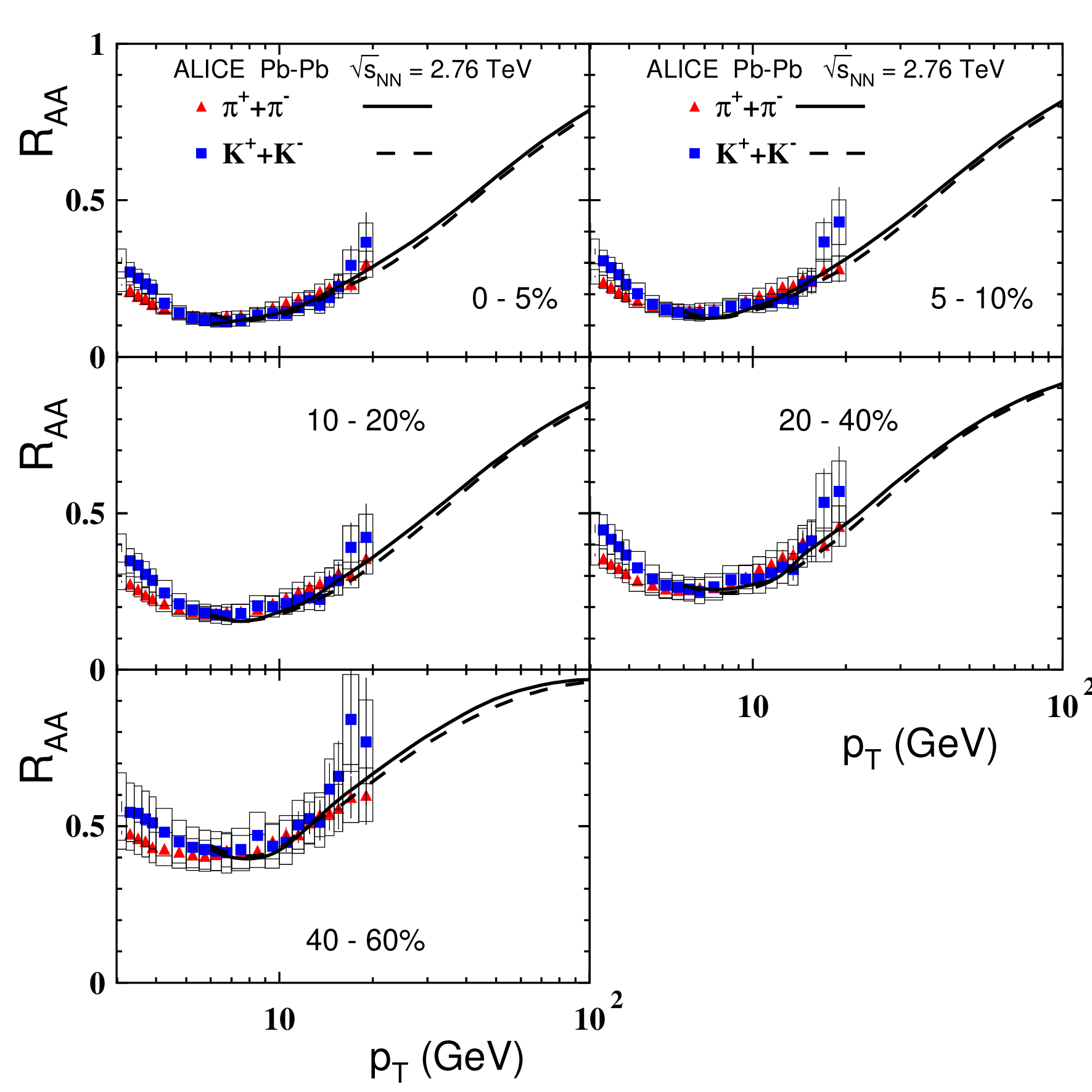}
    \caption{
     %====================
     \label{pions-kaons}
     %====================  
     (color online)
     Model predictions for
     the suppression factor $R_{AA}(p_T)$ in
     inclusive production of pions (solid lines) and kaons 
     (dashed lines) at different centralities
     in lead-lead collisions at $\sqrt{s}_{NN}=2.76\,\TeV$.
     %The suppression factor $R_{AA}(p_T)$ for
     %the prompt $J/\Psi$ at the centrality $0-5\%$
     %in Pb-Pb collisions at $\sqrt{s}_{NN}=5.02\,\TeV$.
     Data for $R_{AA}$ are from the ALICE experiment
     \cite{alice-u16}.
     %and ATLAS \cite{atlas18-prompt-psi} 
     }         
\end{figure}

We performed calculations with constituent quark masses $m_q = 0.3\,\GeV$ 
and $m_s = 0.5\,\GeV$.
The results for the nuclear modification factor $R_{AA}(p_T)$ are compared 
with data on production of pions and kaons at the collision energy 
$\sqrt{s}_{NN} = 2.76\,\TeV$ and at different centralities in Fig.~\ref{pions-kaons}. 
As mentioned above, the rate of broadening, parameter $\hat q_0$, 
could not be predicted reliably because of poorly known non-perturbative dynamics. 
So it was treated as a free parameter and adjusted to data at 
$\hat q_0=2.0\GeV^2/\fm$ \cite{green}.
We assume that $\hat q_0$ is the same for pions and kaons, 
only the mean radii are different. For heavy flavors the values of $\hat q_0$ 
turn out to be quite less (see below).

\begin{figure}[hbt]
    \includegraphics[height=8.9cm]{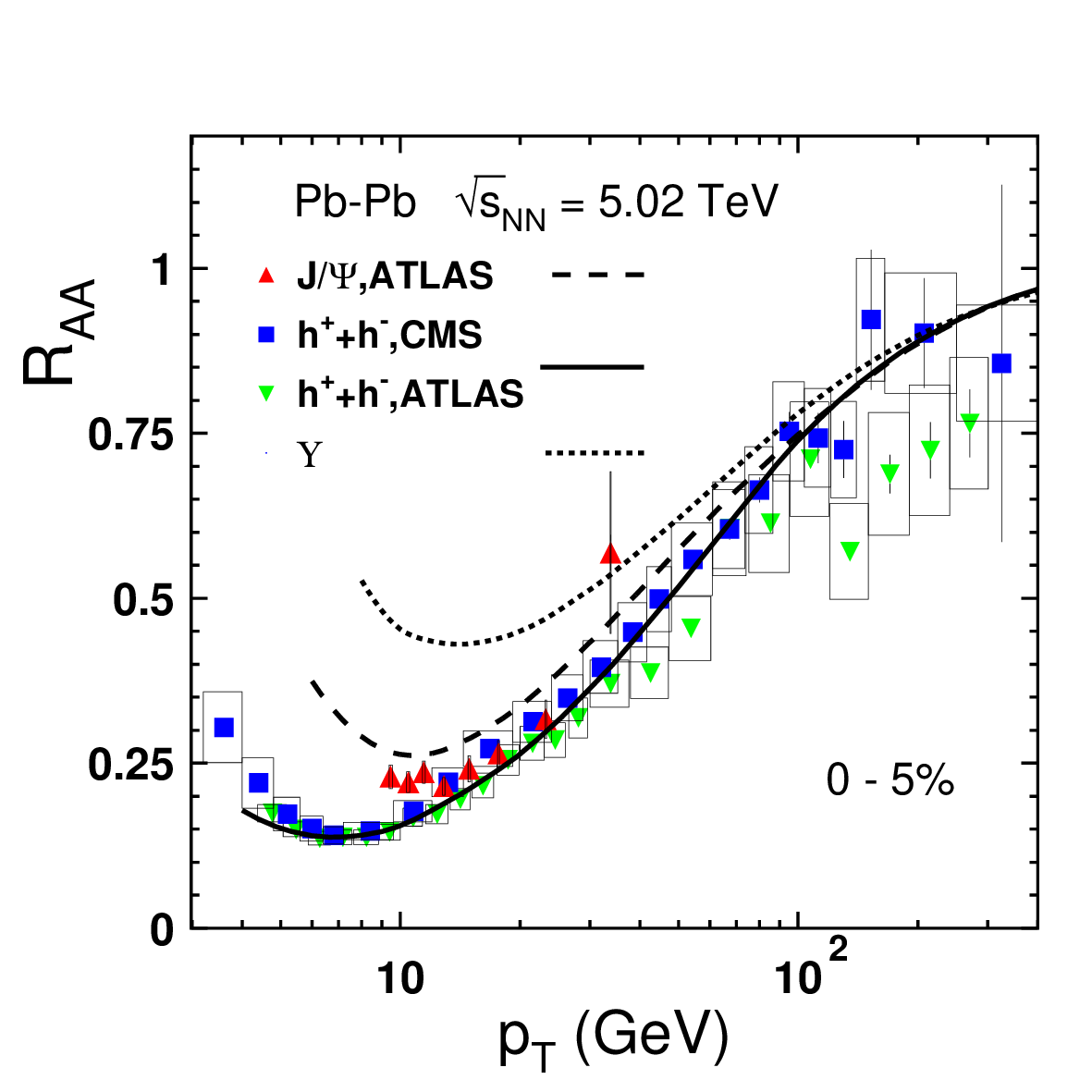}
    \caption{ 
    %=====================
    \label{hadrons-psi-y}
    %=====================
    (color online)
    Model predictions for
    the suppression factor $R_{AA}(p_T)$ in
    inclusive production of charged hadrons (solid line), prompt $J/\Psi$
    (dashed line) and $\Upsilon$ (dotted line)
    at the centrality $0-5\%$ in lead-lead collisions  
    at $\sqrt{s}_{NN} = 5.02\,\TeV$.
    Data for $R_{AA}$ in production
    of charged hadrons are from the
    ATLAS \cite{atlas17-h} and CMS \cite{cms16-h} collaborations.
    Data for $R_{AA}$ in production of prompt $J/\Psi$ are
    from the ATLAS \cite{atlas18-prompt-psi} measurements.
    }         
\end{figure}

With the same value of $\hat q_0$ for light quarks we also describe well data 
for the magnitude of suppression for charged light hadrons h$^{\pm}$ produced 
in central HICs at the collision energy $\sqrt{s}_{NN} = 5.02\,\TeV$, depicted in
Fig.~\ref{hadrons-psi-y}.

%...................... FIG.2 ..............................
%%%%%%%%%%%%%%%%%%%%%%%%%%%%%%%%%%%%%%%%%%%%%%%%%%%%%%%%%%%%
%%%%%%%%%%%%%%%%%%%%%%%%%%%%%%%%%%%%%%%%%%%%%%%%%%%%%%%%%%

%..................... FIG.3 .............................
%%%%%%%%%%%%%%%%%%%%%%%%%%%%%%%%%%%%%%%%%%%%%%%%%%%%%%%%%%
%%%%%%%%%%%%%%%%%%%%%%%%%%%%%%%%%%%%%%%%%%%%%%%%%%%%%%%%%%

One can see from Figs.~\ref{pions-kaons} and \ref{hadrons-psi-y} 
that pions, kaons and charged hadrons
(combination of pions, kaons and protons with a dominant weight of pions)
are suppressed similarly, which is not a big surprise due to similarity of the
color transparency effects for all particles, controlled by the formation length. 
The latter is the path length required for the dipole to expand up to hadronic radius,
%
%------------------------------------------------------------------------
\beq
L_f \approx 
\frac{1}{8}\,R_{h}^2\,\sqrt{p_T^2 + 4 m_q^2}
=
\frac{1}{3}\,\la r_{ch}^2\ra_h\,\sqrt{p_T^2 + 4 m_q^2}\,.
%==============
\label{lf-sym}
%==============
\eeq
%------------------------------------------------------------------------
%
Here we simplify the dipole evolution, by treating it as a classical expansion.
Therefore, the results should be taken with caution, only as semi-qualitative estimate.

According to Eq.~(\ref{lf-sym}) the color transparency regime becomes effective above
the following $p_T$-values,
%
%-------------------------------------------------------------------
\beq
E_T
\approx 
p_T
\gsim 
\frac{3\,R_A}{\la r_{ch}^2\ra_h} 
=
\frac{8\,R_A}{R_{h}^2}
= \left. p_T\right|_{min}^h\,.
%\label{207}
\eeq
%-------------------------------------------------------------------
%
Taking in Eq.~(\ref{lf-sym}) $L_f \approx R_A$ with the nuclear radius $R_A\approx 5\fm$, 
and $\la r_{ch}^2\ra_{\pi} = 0.659^2\,\fm^2$,
$\la r_{ch}^2\ra_K = 0.560^2\,\fm^2$
(see compilations of experimental measurements
on \url{https://pdglive.lbl.gov/DataBlock.action?node=S008CR} and \url{https://pdglive.lbl.gov/DataBlock.action?node=S010CR}),
we obtain the following minimum values of $p_T$ for the color transparency regime,
%
%-------------------------------------------
\beqn
\left. p_T\right|_{min}^{\pi}&\sim& 8\GeV,
\nonumber\\
\left. p_T\right|_{min}^K&\sim& 10\GeV\,.
%\label{212}
\eeqn
%-------------------------------------------
%
Such an expectation is consistent with the ALICE data 
and model calculations presented in Fig.~\ref{pions-kaons}.
We will demonstrate and discussed below that a similar suppression 
of other large-$p_T$ hadrons is less obvious.

%----------------------------------------------------
%---------------------------------------------------
\subsection{Heavy flavored mesons produced in HICs
%===========
\label{hfm}
%===========
}
%---------------------------------------------------
%----------------------------------------------------

Mesons with open heavy flavor, such as $D$- and $B$-mesons, represent 
an asymmetric heavy-light systems. 
The transverse separation of the $Q\bar q$ dipole is controlled 
by the mass of the light quark, $r\sim 1/m_q$, while the longitudinal 
momentum is nearly entirely carried by the heavy quark. 
The light $\bar q$ carries only a small fraction $\alpha\sim m_q/m_Q$ 
of the dipole LF momentum. 
Such a specifics of heavy-light mesons leads to the following key features 
of their production and final-state interactions: 
\\
%%%%%

(i) 
%%%%%
As far as the heavy quark carries the main fraction of the LF momentum 
of the meson, the fragmentation function $D_{Q/Q\bar q}(z)$ should peak 
at large $z$, as is confirmed by data on $e^+e^-$ annihilation 
\cite{charm-bottom1,charm-bottom2};
\\
%%%%%

(ii) 
%%%%%
Smallness of 
%==============
the LF momentum fraction
%==============
$1-z$ indicates at a short radiational path length. i.e. fragmentation 
of a heavy quark ceases promptly, on a path length much shorter than 1 fm 
\cite{hm17,hm18,universal19,hq19a,hq19b,hm23}. 
Therefore the early perturbative stage of hadronization and long-range final 
state interaction in a nuclear environment factorize and can be treated separately
\cite{hm17,hm18,universal19,hq19a,hq19b,hm23};
\\
%%%%%%

(iii) 
%%%%%%
An asymmetric $Q\bar q$ dipole, produced with a small initial separation, 
expands quickly up to the meson size $\la r_M^2\ra$, related to the mean 
charge radius squared $\la r_{ch}^2\ra_M$ of the produced heavy 
(either $D$ or $B$) mesons, $\la r_M^2\ra = 8/3\la r_{ch}^2\ra_M$.
The transverse speed of expansion is inverse to the quark mass, 
i.e. is dominated by the light component of the dipole. 
At the same time, the Lorentz factor is controlled by the heavy quark mass. 
Thus, the formation length in the rest frame of the medium reads,
%
%-----------------------------------------------------
\beq
%t_f\approx \frac{m_q\,\sqrt{p_T^2+m_Q^2}}{2m_Q}\,
%\la r_M^2\ra
%\Rightarrow
L_f\approx \frac{4m_q\,p_T}{3m_Q}\,
\la r_{ch}^2\ra_M\,,
%===============
\label{lf-asym}
%===============
\eeq
%-----------------------------------------------------
%
demonstrating thus a significant reduction in comparison with a light-light dipole, 
Eq.~(\ref{lf-sym});
\\
%%%%%

(iv) 
%%%%%
A $Q\bar q$ dipole, expanded quickly its size up to a large value 
$\la r_{ch}^2\ra_M \approx \la r_{ch}^2\ra_{\pi}$, 
has a large break-up inelastic cross section, i.e. a very short mean free path, 
$\lambda_{Q\bar q} = 1/\hat q \la r_M^2\ra\ll 1\,\fm$, in a dense medium. 
Nevertheless, a break-up of such a dipole does not produce a dramatic effect, 
since the heavy quark carrying the main momentum fraction, 
easily picks up a light $\bar q$ and produces a new dipole 
with about the same momentum.
%
%Nevertheless, 
However,
after many such breaks on the long path in the medium, the leading heavy quark
may lose a substantial part of its momentum. 
This leads to a positive shift $\Delta z$ in the fragmentation function 
$D_{Q/Q\bar q}$ resulting in suppression of the production rate.

\begin{figure}[hbt]
    \includegraphics[height=8.9cm]{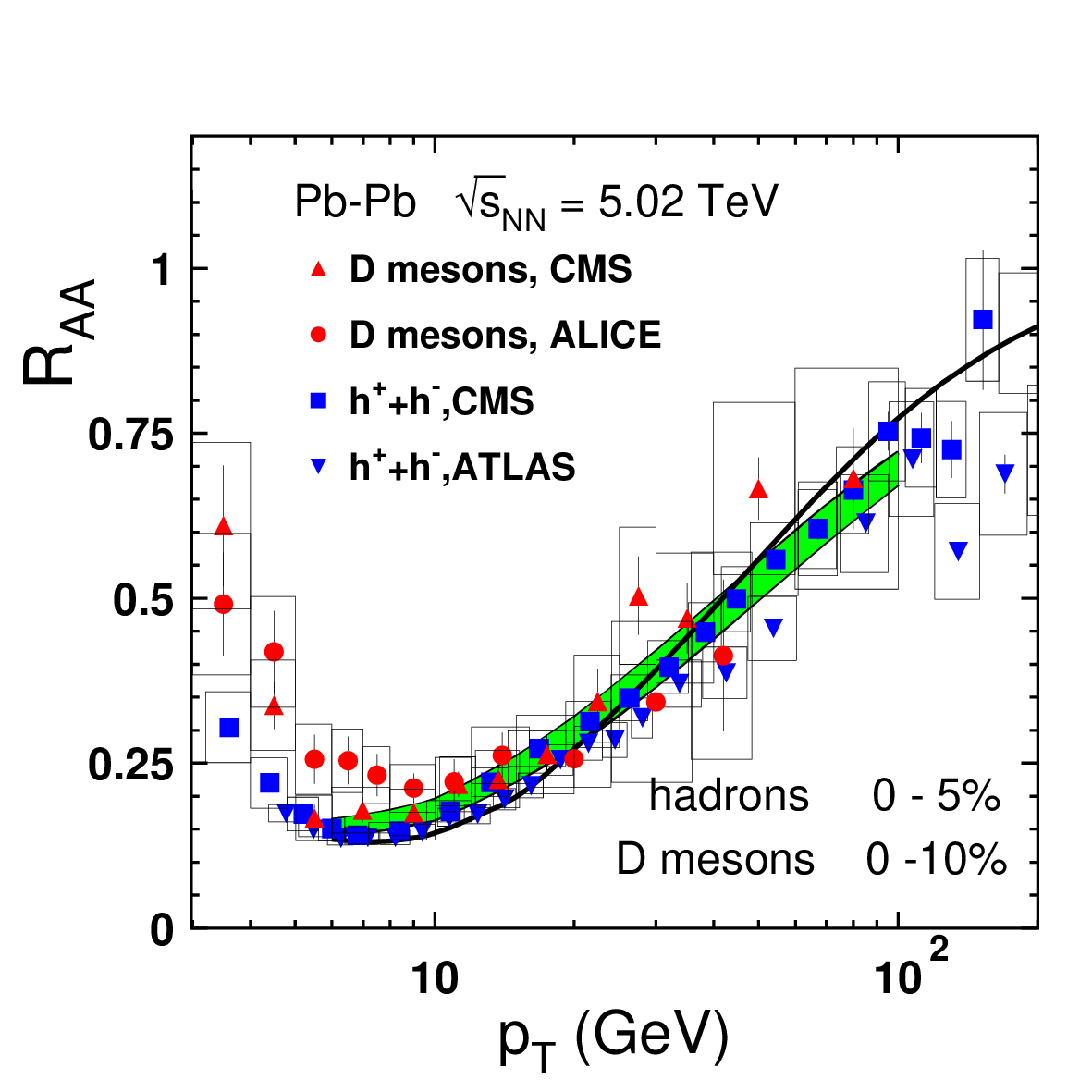}
    \caption{ 
    %===================
    \label{hadrons-d} 
    %===================
    (color online)
    Comparison of model predictions with data for
    the suppression factor $R_{AA}(p_T)$ in
    inclusive production of charged hadrons (solid line) and
    heavy flavored $D$ mesons (green strip)
    at the centrality $0-5\%$ and  $0-10\%$, respectively.
    Data on $R_{PbPb}(p_T)$ at $\sqrt{s}_{NN} = 5.02\,\TeV$
    in production of charged hadrons are from the
    ATLAS \cite{atlas17-h} and CMS \cite{cms16-h} collaborations.
    %
    %Data for production of $B$ mesons are
    %from the ATLAS \cite{b-atlas} and CMS \cite{cms-B-pp} measurements.
    %
    Data for production of $D$ mesons are
    from the ALICE \cite{D-alice} and CMS \cite{cms-D-pp} experiments.
    }
\end{figure}

\begin{figure}[hbt]
    \includegraphics[height=8.9cm]{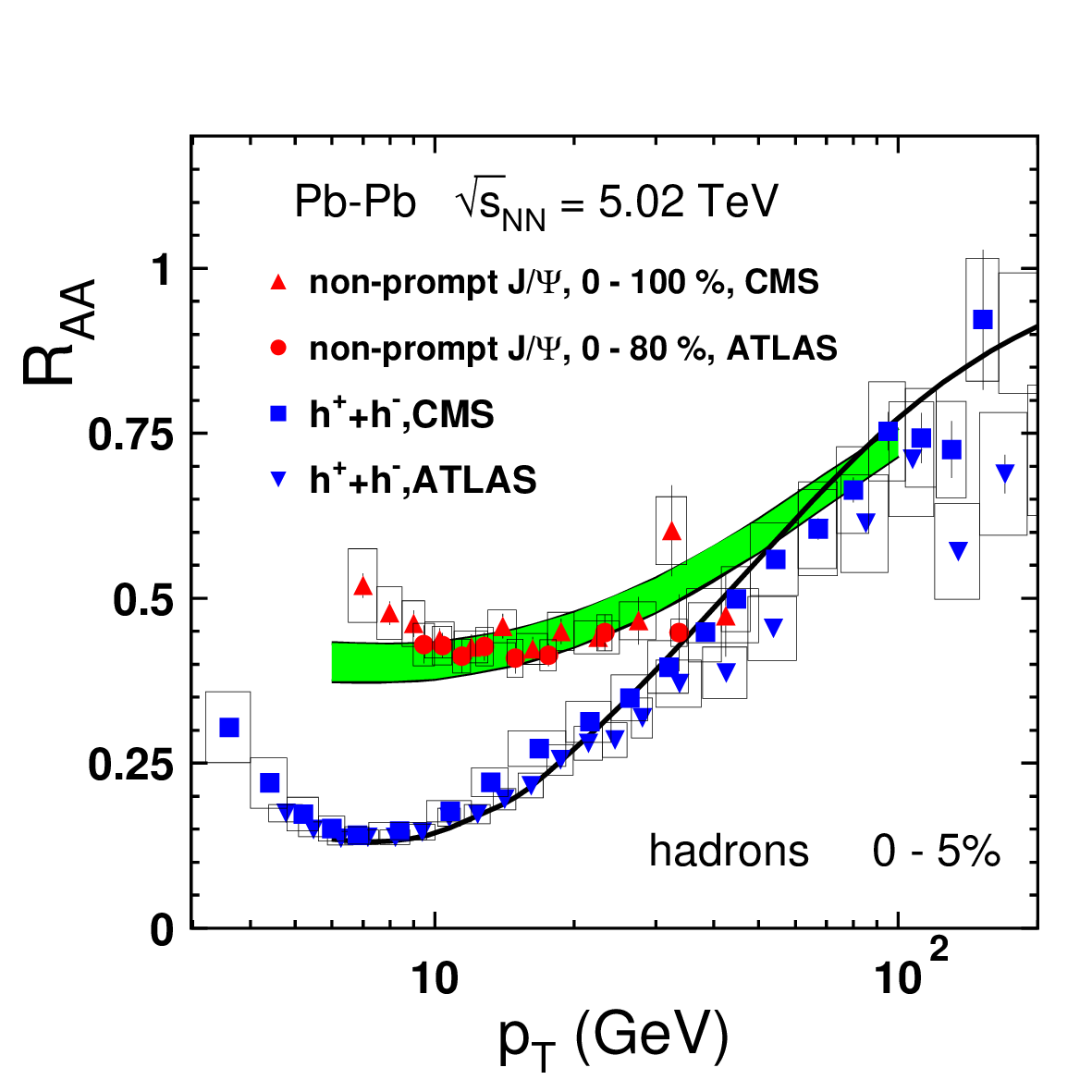}
    \caption{ 
    %===================
    \label{hadrons-b} 
    %===================
    (color online)
    The same as Fig.~\ref{hadrons-d} but now
    the suppression factor $R_{AA}(p_T)$ for
    inclusive production of charged hadrons 
    at the centrality $0-5\%$
    is compared to
    heavy flavored $B$ mesons produced at all centralities.
    %Data on $R_{PbPb}(p_T$ at $\sqrt{s}_{NN} = 5.02\,\TeV$
    %in production of charged hadrons are from the
    %ATLAS \cite{atlas17-h} and CMS \cite{cms16-h} collaborations.
    %
    Data for production of $B$ mesons are
    from the ATLAS \cite{b-atlas} and CMS \cite{cms-B-pp} 
    experiments.
    %measurements.
    %
    %Data for production of $D$ mesons are
    %from the ALICE \cite{D-alice} and CMS \cite{cms-D-pp} experiments.
    }
\end{figure}

More details of this process can be found in Refs.~\cite{hm17,hm18,universal19,hq19a,hq19b,hm23}. 
The results for the nuclear ratio $R_{AA}(p_T)$ in production of 
$D$ and $B$ mesons are compared with suppression of light charged hadrons 
in Figs.~\ref{hadrons-d} and \ref{hadrons-b}. 

%............................ FIG.4 ......................................
%%%%%%%%%%%%%%%%%%%%%%%%%%%%%%%%%%%%%%%%%%%%%%%%%%%%%%%%%%%%%%%%%%%%%%%%%
%%%%%%%%%%%%%%%%%%%%%%%%%%%%%%%%%%%%%%%%%%%%%%%%%%%%%%%%%%%%%%%%%%%%%%%%%

%............................ FIG.5 ......................................
%%%%%%%%%%%%%%%%%%%%%%%%%%%%%%%%%%%%%%%%%%%%%%%%%%%%%%%%%%%%%%%%%%%%%%%%%
%%%%%%%%%%%%%%%%%%%%%%%%%%%%%%%%%%%%%%%%%%%%%%%%%%%%%%%%%%%%%%%%%%%%%%%%%

In the limit of short formation length in comparison with 
the nuclear radius, $L_f\ll R_A$, no color transparency effects 
causing the rise of $R_{AA}$ vs $p_T$ are expected. 
However, the formation length $L_f\propto E_T$ rises with $E_T$ 
leading so to onset of the color transparency regime. This happens at
the following value of $p_T$,
%
%-----------------------------------------------------------------------
\beq
p_T
\gsim 
\frac{2m_Q\,R_A}{m_q\la r_M^2\ra} 
= 
\left. p_T\right|_{min}^M
\approx
\left. p_T\right|_{min}^h\cdot \frac{m_Q}{4 m_q}
\,.
%===========
\label{207}
%===========
\eeq
%----------------------------------------------------------------------
%
Setting $R_A\approx 5\fm$, and $\la r_M^2\ra \sim 1/m_q^2$ we obtain
the minimum values of $p_T$ for the color transparency regime,
%
%-----------------------------------------------
\beqn
\left. p_T\right|_{min}^D&\sim& 15\div 20\GeV,
\nonumber\\
\left. p_T\right|_{min}^B&\sim& 45\div 50\GeV\,.
%===========
\label{212}
%===========
\eeqn
%-----------------------------------------------
%
This estimate is confirmed by more accurate calculations \cite{hm17,hm18,hq19b,hm23} 
and by data displayed in Figs.~\ref{hadrons-d} and \ref{hadrons-b}. 
Indeed, the data in Fig.~\ref{hadrons-d} show a steep rise of $R_{AA}$ 
for $D$-mesons starting from $p_T\sim 15\div 20\GeV$.
Data on $R_{AA}$ for $B$-mesons indeed demonstrate a flat behavior up to 
$p_T\approx 40\GeV$, and we expect onset of the rising $R_{AA}(p_T)$ 
%%%%%
as
%%%%%
for light hadrons at $p_T\gsim 50\GeV$ in agreement with the 
above model calculations \cite{hm23} presented in Fig.~\ref{hadrons-b}.

The important feature of our results is a universal suppression 
for production of different mesons. 
The corresponding broadening rate parameter (transport coefficient) 
was found to be $\hat q_0^c=0.45-0.55 \GeV^2/\fm$ for $D$-mesons, and $\hat q_0^b=0.2-0.25 \GeV^2/\fm$ for $B$-mesons (shown by strips in Figs.~\ref{hadrons-d} and \ref{hadrons-b}).
The value of $\hat q_0^c$ for the Large-Hadron-Collider (LHC) kinematic region is rather consistent
with $\hat q_0^c=0.3-0.5 \GeV^2/\fm$ found in our previous works \cite{q0-1,q0-2} 
analyzing charmonium suppression in gold-gold collisions at Relativistic-Heavy-Ion-Collider (RHIC).

%-------------------------------------------------------
%-------------------------------------------------------
\subsection{Heavy quarkonia produced in HICs
%===============
\label{hq-hics}
%===============
}
%-------------------------------------------------------
%-------------------------------------------------------

Here we highlight briefly the similarities and differences in production and final-state interactions
of heavy quarkonia (HQ) compared to light and heavy flavored mesons:
\\
%%%%%

(i)
%%%%%
As was demonstrated in Sec.~\ref{lp-vacuum} and showed in Fig.~\ref{fig:eloss-qn}, 
the light and heavy quarks radiate differently. 
The duration of the fragmentation process decreases with the quark mass and the 
corresponding path length $L_p\ll 1\,\fm$, similarly as in production of heavy flavored mesons; 
\\
%%%%%

(ii)
%%%%%
Smallness of the fractional energy radiated by heavy quarks leads to 
a specific shape of the fragmentation functions (FFs)
$c\to J/\Psi$ and $b\to \Upsilon$ consistent with calculations
\cite{ffc-nlo} including the next-to-leading-order corrections.
Similar shapes of FFs $c\to D$ and $b\to B$ were predicted in
our recent work \cite{hm23} in a good accord with direct measurements 
in $e^+e^-$ annihilation \cite{charm-bottom1,charm-bottom2}.
The corresponding distributions strongly peak at $z\sim0.75\div 0.85$ 
due to almost the whole momentum of the jet carried by
the final heavy flavored mesons and/or quarkonia.
On the contrary, the FFs of light quarks to light mesons fall steadily 
and steeply from small $z$ towards $z=1$ \cite{kkp};
\\
%%%%%%

(iii)
%%%%%%
In contrast to heavy flavored mesons, the HQ are produced mainly
from the symmetric $Q\bar Q$ dipoles with $\alpha\sim 0.5$.
The transverse size of the $Q\bar Q$ dipole is controlled by the mass of the 
heavy quark, $r\sim 1/m_Q$, differently from production of heavy flavored mesons. 
Consequently, expansion of such a dipole in the medium with small initial separation, 
$r_0\sim 1/E_T$, up to formation of the final small-sized heavy quarkonia, is much longer.
However, it is slowed down due to Lorentz time dilation, so is subject to  
the effect of  color transparency. Therefore, the mechanism of final-state interactions 
is different from that in production of heavy flavored $D$ and $B$ mesons discussed 
in the previous Sec.~\ref{hfm}; 
\\
%%%%%%

(iv)
%%%%%%
According to Eq.~(\ref{lf-sym}) the quarkonium formation time, 
controlling the onset of color transparency, is longer for heavier quarks.
We expect a hierarchy, $L_f(\Upsilon) < L_f(J/\psi) < L_f(\pi,K,p)$, 
related to the inequalities between average meson sizes, 
$\la r_{\Upsilon}^2\ra < \la r_{J/\psi}^2\ra < \la r_{\pi,K,p}^2\ra$.
Consequently, at smaller values of $p_T\lsim 10\div 15\,\GeV$, 
the in-medium final state interaction in production is controlled 
by the stage of fast expansion of corresponding dipoles demonstrating 
a sensitivity to quark masses. 
It is manifested in Fig.~\ref{hadrons-psi-y} as a hierarchy, 
$R_{AA}(h^++h^-) < R_{AA}(J/\psi) < R_{AA}(\Upsilon)$.

For a more accurate description of the color transparency regime, 
an alternative to the classical expansion Eq.~(\ref{lf-sym}), 
more appropriate is the quantum-mechanical expression for the formation time,
%
%--------------------------------------------
%\beq
%t_f = \frac{2\,E_T}{{m'}_{HQ}^2 - m_{HQ}^2}\,,
%=============
%\label{tf-hq}
%=============
%\eeq
%---------------------------------------------
%
%with the corresponding coherence length given as,
%
%----------------------------------------------
\beq
L_f
%t_f \cdot \sqrt{1 - \frac{ 4m_Q^2}{E_T^2}}
\approx
\frac{2\,p_T}{{m'}_{HQ}^2 - m_{HQ}^2}\,,
%=============
\label{lf-hq}
%=============
\eeq
%-----------------------------------------------
%
where $m_{HQ}$ and ${m'}_{HQ}$ is the mass of the ground state and next
excited state of heavy quarkonia, respectively.
Eq.~(\ref{lf-hq}) is based on the uncertainty principle. 
One can distinguish between production of these heavy quarkonium states 
only if the process lasts sufficiently long compared with the inverse 
difference of corresponding masses. 
Besides, Eq.~(\ref{lf-hq}) includes the Lorentz time dilation 
in the rest frame of the nucleus.
So the full CT regime is reached for sufficiently quarkonium transverse momenta,
%
%-------------------------------------------------------------
\beq
p_T\gsim 
%\frac{8\,R_A}{\la r_M^2\ra} = \left. p_T\right|_{min}^M\,,
\frac{1}{2}\,R_A\,({m'}_{HQ}^2 - m_{HQ}^2)
= \left. p_T\right|_{min}^{HQ}
\approx
\left. p_T\right|_{min}^{h}\cdot \frac{m_Q}{m_q}
\,, 
%===========
\label{407}
%===========
\eeq
%-------------------------------------------------------------
%
giving the following minimum values of $p_T$ for the CT regime,
%
%----------------------------------------------------
\beqn
\left. p_T\right|_{min}^{J/\psi}&\sim& 50\GeV,
\nonumber\\
\left. p_T\right|_{min}^{\Upsilon}&\sim& 140\GeV\,.
%===========
\label{412}
%===========
\eeqn
%---------------------------------------------------
%
These evaluations are confirmed by our model calculations with results 
depicted  in Fig.~\ref{hadrons-psi-y}. 
Here the data for charmonium production demonstrate that the suppression factor 
$R_{AA}(p_T)$ is similar to that for light hadrons in correspondence with our predictions. 
The data for the large-$p_T$ bottomonium production in HICs are plotted for comparison, 
to show an evidence of universal behavior of the corresponding $R_{AA}$ at large $p_T$ 
given mainly by the CT effects.

Similarly as in production of $D$ and $B$ mesons, we have obtained 
the model predictions, demonstrating the universal suppression also 
in production of different heavy quarkonia, adjusting the maximum value of
the broadening rate in consistency with Ref.~\cite{hm23}, where $\hat q_0^c=0.45-0.55\,\GeV^2/\fm$ for $J/\psi$-mesons, and $\hat q_0^b=0.2-0.25\,\GeV^2/\fm$ for $\Upsilon$-mesons.

%%%%%%%%%%%%%%%%%%%%%%%%%%%
\section{Conclusions
%====================
\label{conclusions}
%====================
}
%%%%%%%%%%%%%%%%%%%%%%%%%%%

In the present paper we analyzed the nontrivial observation of 
similar magnitudes of the nuclear suppression factor $R_{AA}(p_T)$ 
for production of different hadrons at large $p_T$. 
The main conclusions are as follows:

\begin{itemize}

\item 
In the process of high-$p_T$ meson production with Feynman $x_F=0$ in HICs 
one can single out two stages. 
The first one is accompanied by the gluon radiation and parton energy loss 
up to color neutralization and ceases shortly after a hard collision. 
Radiative energy loss at this stage is nearly the same in vacuum and nuclear medium. 
Remarkably, heavy quarks radiate a much smaller fraction of the initial 
quark energy compared to light quarks due to the dead-cone effect. 
This is why heavy flavored mesons usually carry a dominant fraction $z$ 
of the heavy quark momentum, in consistency with a non-monotonic shape 
of $z$-dependence of the corresponding fragmentation function. 
In contrast, FFs for light quarks exhibit a monotonically falling $z$-behavior.
The production length $L_p$ of color neutralization, controlling the duration 
of this hadronization stage, turns out to be very short, $L_p\ll 1\div 2\,\fm$ 
for the both light and heavy quarks, and it decreases with $p_T$.

\item 
The first stage does not produce any sizeable suppression effect, 
it is quite short and ends up with production of a small size $\sim 1/p_T$ dipole. 
During subsequent propagation of this dipole throughout the medium
it is developing the wave function of the final meson. 
This may take a long time. The corresponding path length $L_f$ rises with $p_T$. 
This stage of meson production is the main source of suppression, related to 
a chance that the colorless dipole can be broken by inelastic (color-exchange) 
collisions within the medium.

\item 
At large $p_T$, the evolution of a small dipole propagating through a medium 
up to formation of the final meson is slowed down due to Lorentz dilation. 
If $L_f\gsim R_A$, production of various mesons is governed by the color 
transparency effects, and their attenuation is directly controlled by the 
$p_T$-dependent dipole size. 
This leads to independence of the meson radius, i.e. to universality of the 
suppression in production of various high-$p_T$ mesons, analyzed in the present paper.

\item 
The condition $L_f\gsim R_A$ requires sufficiently high $p_T$, 
i.e. imposes the lower limits $p_T|_{min}^M$  defining the range of $p_T$-values, 
$p_T\gsim p_T|_{min}^M$, for the onset of the color transparency regime.
The corresponding  $p_T|_{min}^M$ have been evaluated for pions, kaons, heavy quarkonia, 
as well as for heavy flavored $D$ and $B$ mesons taking into account 
a different dynamics of their production.

\item 
Whereas light mesons (pions, kaons) require a small  $p_T|_{min}^M\sim 8\div 10\,\GeV$
for manifestation of universal suppression, heavy quarkonia prefers higher values,
$p_T|_{min}^{J/\psi}\approx p_T|_{min}^M\cdot f_c\sim 50\,\GeV$ and 
$p_T|_{min}^{\Upsilon}\approx p_T|_{min}^M\cdot f_b\sim 140\,\GeV$,
enlarged by factors $f_c \approx m_c/m_q$ and $f_b \approx m_b/m_q$, respectively.
On the other hand, the heavy flavored $D$ and $B$ mesons scan the smaller 
$p_T|_{min}^{D}\approx p_T|_{min}^M\cdot h_c\sim 15\div 20\,\GeV$ and 
$p_T|_{min}^{B}\approx p_T|_{min}^M\cdot h_b\sim 45\div 50\,\GeV$
due to their asymmetric heavy-light quark configurations and different mechanism of
production leading to smaller factors $h_c \approx m_c/4 m_q < f_c$ and $h_b \approx m_b/4 m_q < f_b$.

\item 
Our analysis of the universal suppression contains one 
unavoidable parameter, which is the broadening rate of the quark in the medium. 
In correspondence with our previous studies, we found its maximal value 
$\hat q_0\sim 2.0\,\GeV^2/fm$, $0.45 - 0.55\,\GeV^2/fm$ and $0.20 - 0.25\,\GeV^2/fm$
for the light, c and b quark, respectively.
Such a diminution of $\hat q_0$ with the quark mass is a clear manifestation of 
the dead-cone effect, which reduces the broadening.

\item 
Our expectations of the manifestation of universal suppression in production of various
high-$p_T$ mesons are consistent with available data, as well as with our model calculations
based on the rigorous Green function formalism.

\vspace*{0.1cm}

\end{itemize}
%%%%%%%%%%%%%%%%%%%%%%%%%%%%%%%%%%%%%%%%%%%%%%%%%%%%

\begin{acknowledgments}

This work was supported in part by grants ANID-Chile FONDECYT 1231062, ANID PIA/APOYO AFB220004, 
and ANID-Millennium Science Initiative Program ICN2019\_044.
The work of J.N. was partially supported 
by the Slovak Funding Agency, Grant No. 2/0020/22.

\end{acknowledgments}

%%%%%%%%%%%%%%%%%%%%%%%%%%%%%%%%%%%%%%%%%%%%%%%%%%%%

\end{document}